\documentclass[aps,twocolumn,showpacs,pra]{revtex4-1}
\usepackage{graphicx,epstopdf}
\usepackage{amsmath}
\usepackage{amssymb}
\usepackage{color}
\usepackage{float}
\usepackage{epsfig}
\usepackage{epstopdf}
\usepackage{palatino}
\usepackage{enumitem}
\usepackage{tcolorbox}
\usepackage{bmpsize}

\usepackage[colorlinks=true, citecolor=blue, urlcolor=blue ]{hyperref}
\input{epsf}
\begin{document}

\title{Fibonacci sequence and its generalizations in doped quantum spin ladders}

\author{Sudipto Singha Roy,\(^{1,2,3}\) Himadri Shekhar Dhar,\(^{4}\) Aditi Sen(De),\(^{3}\) and Ujjwal Sen\(^{3}\)}

\affiliation{\(^1\)Instituto de F{\'i}sica T{\'e}orica  UAM/CSIC,  Madrid,  Spain}
\affiliation{\(^2\)Department of Applied Mathematics, Hanyang University (ERICA), 55 Hanyangdaehak-ro, Ansan, Gyeonggi-do, 426-791, Korea}
\affiliation{\(^3\) Harish-Chandra Research Institute, HBNI, Chhatnag Road, Jhunsi, Allahabad 211 019, India}
\affiliation{\(^4\)Institute for Theoretical Physics, Vienna University of Technology, Wiedner Hauptstra{\ss}e 8-10/136, A-1040 Vienna, Austria}

\date{\today}

\begin{abstract}
An interesting aspect of antiferromagnetic quantum spin ladders, with complete dimer coverings, is that the wave function can be recursively generated by estimating the number of coverings in the valence bond basis, which follow the fabled \emph{Fibonacci sequence}. In this work, we derive generalized forms of this sequence for multi-legged and doped quantum spin ladders, which allow the corresponding dimer-covered state to be recursively generated. We show that these sequences allow for estimation of physically and information-theoretically relevant quantities in large spin lattices without resorting to complex numerical methods. We apply the formalism to calculate the valence bond entanglement entropy, which is an important figure of merit for studying cooperative phenomena in quantum spin systems with SU(2) symmetry. We show that introduction of doping may mitigate, within the quarters of entanglement entropy, the dichotomy between odd- and even- legged quantum spin ladders.
\end{abstract}

\maketitle

\section{Introduction}

%

\begin{figure}[h]
\begin{center}
\includegraphics[width=2.6in,angle=00]{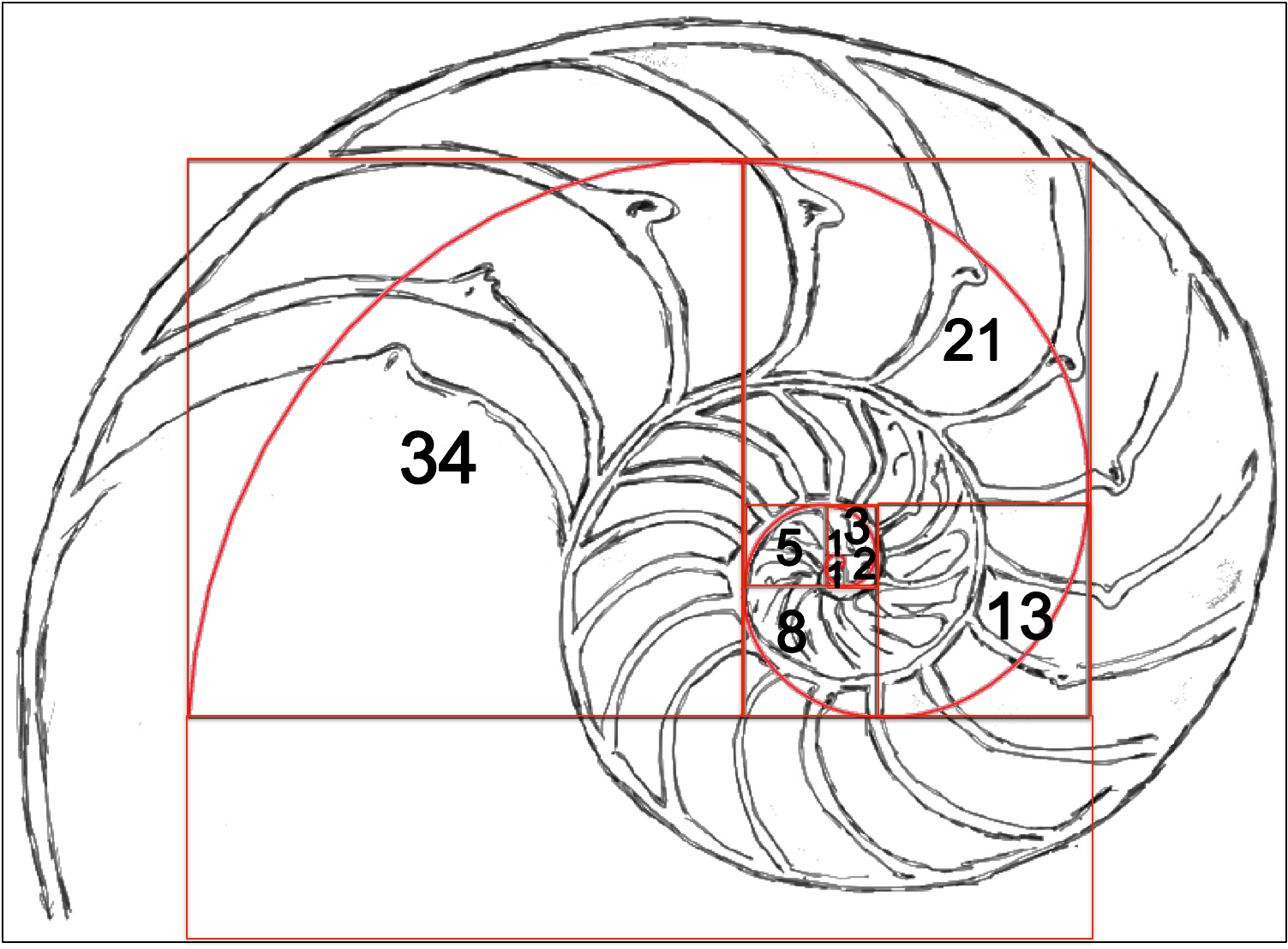}
\end{center}
\caption{(Color online.) The naturally occurring patterns on the molluscan shell roughly follows the golden spiral, with the areas of the square blocks following the Fibonacci sequence.}
\label{molluscan}
\end{figure}

An important class of many-body systems, in contemporary research on the spin-singlet ground states of the spin-1/2 antiferromagnetic (AFM) Heisenberg model, 
is the resonating valence bond (RVB) states \cite{andersonRVB}, which were introduced to study Mott-insulators \cite{andersonMott} and high-T$_c$ superconductivity in cuprates \cite{andersonSup,roksharSup,bhaskaranRVB}. Over the years, RVB states have been significant in investigation of exotic quantum phenomena ranging from spin liquids, topological phases of matter \cite{RVBsl,RVBsl2,RVBtopo}, and quantum correlations \cite{delgadoRecur1,delgadoRecur3,fanRecur,delgadoRecur2,kim,RVBenttopo,RVBent,hsdRecur1,hsdRecur2,sroyRecur1,sroyRecur2}. 
For low dimensional quantum AFM spin systems the ground state (GS) is often represented in terms of the short-range RVB ansatz \cite{expo1}, which is well supported by various numerical and tensor-network methods \cite{expo1, gopalan,wiese,scala}.
Moreover, there are instances where short-ranged RVB states have been shown to be the exact ground state of important 
strongly-correlated spin models over distinct regions in the parameter space \cite{sroyRecur1,expo1, gopalan,wiese,scala,sroyRecur2,MG,BK,Mila,eggm,tj,doped_sc2}. 
%
An interesting feature of the short-range RVB state, with spin singlets between nearest-neighbor (NN) spins, is that the state for an arbitrary number of spins can be recursively generated from smaller states \cite{delgadoRecur1,delgadoRecur3,fanRecur,delgadoRecur2,kim,hsdRecur1,hsdRecur2,sroyRecur1,sroyRecur2}. 
In particular, for spin models in a ladder configuration with NN dimers the number of possible dimer coverings in the RVB state is given by the \textit{Fibonacci sequence} \cite{fibonacci}. More precisely, the number of coverings in the valence bond basis, for an $N$-rung short-range RVB ladder state, is the sum of the possible coverings for ($N-1$)- and ($N-2$)-rung RVB ladders \cite{delgadoRecur1,fanRecur}. It is often said that the Fibonacci numbers are nature's numbering system. There exist an enormous number of examples where the basic structure of naturally occuring patterns, such as phyllotaxis and flowering in plants, arrangement in pine cones and pineapples, pedigrees in honeybees, and various shell proportions in molluscs (see Fig.~\ref{molluscan}), follow the Fibonacci sequence (cf. Refs.~\cite{fibonacci,goldratio}).

The presence of such a sequence for the dimer coverings  has a direct bearing on analytical computation of important physical quantities in RVB systems. Several relevant system properties, such as two-site correlation functions \cite{delgadoRecur1,fanRecur,delgadoRecur2} and short-range and global entanglement \cite{hsdRecur1,hsdRecur2,sroyRecur1,sroyRecur2} can be obtained from the terms in the sequence, without employing any approximate numerical or mean-field methods even for a large number of lattice sites. 
For instance, in short-range RVB states on a spin ladder, with equal-weight superposition of all possible NN dimer coverings, the valence bond entanglement entropy \cite{aletVBE,wojcikVBE} across any bipartition can be derived as a function of the number of coverings in each contiguous block on either side of the bipartition. These numbers are then obtained directly from the Fibonacci sequence, which for large $N$ increases with the \emph{golden ratio}, $\Phi$ \cite{goldratio}. 


In this work, we look at the possibility of deriving generalized sequences, similar to the Fibonacci sequence but for the more complex RVB states, corresponding to multi-legged and doped quantum spin-1/2 ladders.
Beginning with resonant states that arise from quantum spin Hamiltonians, under the RVB ansatz,
we show that a recursion relation, in terms of the number of coverings in the valence bond basis, can be generated to build larger $N$-rung RVB states from smaller rung states
(cf.~\cite{hsdRecur1,hsdRecur2,sroyRecur1,sroyRecur2,delgadoRecur1,fanRecur,delgadoRecur2,kim}). Importantly, a generalized sequence for the number of coverings in these states can be derived. These sequences predictably deviate from the usual Fibonacci sequence, which only corresponds to the case for RVB states in undoped two-legged spin ladders. Nonetheless, these generalized sequences in multi-legged and doped spin ladders allow us the same potential to compute important physical quantities. In particular, we show that one can calculate the valence bond entanglement entropy for these states, which is a recently introduced  estimator of the characteristics of block entanglement entropy \cite{benEE,horoQE} in quantum spin systems with SU(2) symmetry \cite{aletVBE,wojcikVBE}. This allows us to provide important insights into the dichotomy between odd- and even-legged quantum spin ladders and the strong effects of doping, in terms of the variation of the average number of dimers and valence bond entanglement entropy in large quantum spin ladders.


The paper is arranged as follows. 
In Sec.~\ref{fibonacci_summary}, we begin with the derivation the recursion relation, and the Fibonacci sequence for the number of dimer coverings, in two-legged RVB ladders. 
In Sec.~\ref{multi-legged scaling}, we derive the generalized sequences obtained from the recursion relations for multi-legged quantum spin ladders. 
We introduce doped RVB spin ladders in Sec.~\ref{Doped RVB detail}, and obtain the generalized sequences. We define and analyze the valence bond entanglement entropy for the above systems in Sec.~\ref{VBE}.
We end with a conclusion  in Sec.~\ref{Discussion}.

\section{The Fibonacci sequence in resonating valence bond ladder}
\label{fibonacci_summary}

\begin{figure}[t]
\begin{center}

\includegraphics[width=2.5in,angle=00]{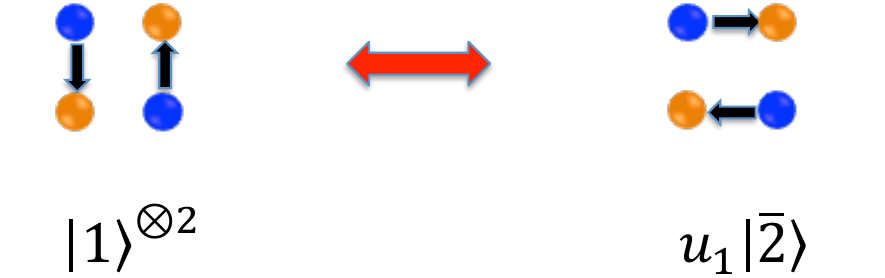}
\caption{(Color online.) The resonance (indicated by two-way arrow) between the dimer coverings in the states $|1\rangle^{\otimes 2}$ generates the configuration $|\bar{2}\rangle$, for the two-legged quantum spin ladders. The sites on different bipartite lattices, $\mathcal{A}$ and $\mathcal{B}$, have been distinguished using two different colors. }
\label{rvb2_terms}
\end{center}
\end{figure}

We begin by taking a look at the nearest-neighbor dimer-covered state, also called the short-range resonating valence bond state, of the two-legged quantum spin ladder. Henceforth, the term RVB in this work, shall imply short-range RVB state, unless stated otherwise. An archetypal Hamiltonian for the two-legged quantum spin-1/2 Heisenberg ladder, supporting an RVB ansatz \cite{delgadoRecur1,delgadoRecur3}, can be written as
\begin{equation}
\mathcal{H}_\textrm{2-leg} = J \sum_{m=1}^{2} \sum_{i=1}^{N-1} \vec{S}^m_i \cdot \vec{S}^m_{i+1} + J' \sum_{i=1}^{N} \vec{S}^1_i \cdot \vec{S}^2_{i},
\label{H2}
\end{equation}
where, $\vec{S}^m_i$ is the spin operator at the $i^{th}$ site on the $m^{th}$ leg, and $J$ ($J'$) is the NN spin-spin interaction along the legs (rungs).
For $J$ = 0, there is no interaction along the legs, and the singlet ground state of the Hamiltonian is the product of dimers or spin-singlets along the rungs of the ladder:  
$|\Psi_{\tilde{N}}\rangle$ = $\frac{1}{\sqrt{2}}(|\uparrow^1_i \downarrow^2_i\rangle-|\downarrow^1_i \uparrow^2_i\rangle)^{\otimes N} \equiv|1\rangle^{\otimes N}$. Here, $\tilde{N} = 2\times N$, is the number of spins in the ladder and open boundary conditions have been considered.
According to the RVB ansatz, as soon as the intra-leg interaction, $J$, is switched on, pairs of NN rung dimers ($|1\rangle^{\otimes 2}$) begin to resonate and generate  pairs of horizontal dimers, $|\bar{2}\rangle$ (see Fig.~\ref{rvb2_terms}), with weightage proportional to the interactions, $u_1 \approx J/J'$ \cite{delgadoRecur3}. Such resonance allows for significant lowering of GS energy. For a ladder with two rungs ($\tilde{N} = 4$), the unnormalized GS is then given by 
$|\Psi_{4}\rangle = |1\rangle^{\otimes 2} + u_1 |\bar{2}\rangle$. All RVB states for the two-legged ladder can be build recursively using the resonant states $|1\rangle$ and  $|\bar{2}\rangle$.

Building on the above ansatz, independent of the lattice or the Hamiltonian, the entire RVB state can be build by defining such dimer states that resonate between NN sites.
We begin by defining a bipartite lattice, with two sublattices, $\mathcal{A}$ and $\mathcal{B}$, where each site on sublattice $\mathcal{A}$ ($\mathcal{B}$) is surrounded by NNs on sublattice $\mathcal{B}$ ($\mathcal{A}$). A dimer or a singlet is formed between any pair of NN spins belonging to different sublattices, and can be written as $[e_k,e_l]=\frac{1}{\sqrt{2}}(|\uparrow_k\downarrow_l\rangle-|\downarrow_k\uparrow_l\rangle)$, such that $k \in \mathcal{A}$ and $l \in \mathcal{B}$. We assume that the singlets are always directed from a site in $\mathcal{A}$ to one in $\mathcal{B}$. A single unique covering of the ladder lattice, with NN dimers, is then given by $|\phi_i\rangle$ = $([e_1,e_2]\otimes [e_3,e_4]\otimes\dots [e_{\tilde{N}-1}e_{\tilde{N}}])_i$, where the odd (even) indices belong to sublattice $\mathcal{A}$ ($\mathcal{B}$), and $\tilde{N}$ is even. All such unique coverings together form an overcomplete valence bond basis. The variational RVB state in this basis, with weights $W_i$, can then be written as
$
|\Psi_\textrm{RVB}\rangle=\sum_{i=1}^{Z} W_i ([e_1,e_2]\otimes [e_3,e_4]\otimes\dots [e_{\tilde{N}-1}e_{\tilde{N}}])_i,
$
where $Z$ is the number of all possible coverings or states in the basis $\{|\phi_i\rangle\}$. With regards to the Hamiltonian for the two-legged ladder (Eq.~(\ref{H2})), the weights $W_i$ would be dependent on the interactions $J$ and $J'$.
We note that the state $|\Psi_\textrm{RVB}\rangle$ is not normalized and importantly, $\{|\phi_i\rangle\}$ is not an orthogonal basis. In general, the complexity of the RVB state is compounded by this lack of orthogonality and the exponential increase of $Z$ with the size of the system. This renders exact methods unfeasible for calculating quantities such as entanglement entropy \cite{benEE,horoQE} in large RVB systems.


Now, returning to the $N$-rung, two-legged spin-1/2 ladder, as defined for the Hamitonian in Eq.~(\ref{H2}), in which all the lattice sites are covered by spins, i.e., the system has no dopants or holes,
%
%
the RVB state $|\Psi_{\tilde{N}}\rangle$ = $|N\rangle$ ($\tilde{N} = 2 \times N$) can be recursively built, using the smaller-rung ladders, $|N-1\rangle$ and $|N-2\rangle$ \cite{delgadoRecur1,fanRecur}. The corresponding recursion relation reads as
\begin{eqnarray}
|N\rangle &=&  |N-1\rangle |1\rangle + u_1|N-2\rangle|\bar{2}\rangle.
\label{2_legged_recursion}
\end{eqnarray}
%
Therefore, the number of dimer coverings, ${Z}_{N}$, in a two-legged, $N$-rung, RVB ladder is given by
\begin{eqnarray}
{Z}_{N}&=&{Z}_{N-1}+{Z}_{N-2}.
\label{2legfibo}
\end{eqnarray}
Now, for $N$ = 1 and 2, the RVB states are $|\Psi_{2}\rangle = |1\rangle$ ($Z_{1}$ = 1) and  $|\Psi_{4}\rangle = |1\rangle^{\otimes 2} + u_1 |\bar{2}\rangle$ ($Z_{2}$ = 2), respectively, which generates the Fibonacci sequence $\{Z_N\}$.  One can set $Z_{0}=1$ without any loss of generality. The first few terms of the sequence are tabulated below: 

\begin{center}
\begin{tabular}{ |p{.65cm}|p{.65cm}|p{.65cm}|p{.65cm}| p{.65cm}|p{.65cm}| p{.65cm}|p{.65cm}| p{.65cm}|p{.65cm}|}
\hline
\multicolumn{10}{|c|}{No. of dimer coverings $Z_{N}$} \\
\hline
$~Z_{1}$&$~Z_{2}$&$~Z_{3}$&$~Z_{4}$&$~Z_{5}$&$~Z_{6}$&$~Z_{7}$&$~Z_{8}$&$~Z_{9}$&$~\dots$\\
\hline
$~~1~$&$~~2~$&$~~3~$&$~~5~$&$~~8~$&$~13~$&$~21~$&$~34~$&$~55~$&$~\dots$\\
\hline
\end{tabular}
\vspace{.2cm}
\end{center}


%
%
An important quantity, from the perspective of general RVB states, is the rate of divergence of the series $\{Z_N\}$ at large $N$.
For convenience of notation, we denote this rate as $\alpha$, although for the Fibonacci sequence it is conventionally denoted as $\Phi$, and referred to as the ``golden ratio''. So,
$\alpha$ = ${Z_N}/{Z_{N-1}}$ and it  quantifies the increase in the number of coverings with increasing rung. For two-legged RVB ladders, $\alpha$  can be derived as
\begin{eqnarray}
Z_{N}=\alpha Z_{N-1}=\alpha^2Z_{N-2},
\label{2-legged_fibonacci}
\end{eqnarray}
for sufficiently large $N$.
Now plugging these terms into  Eq.~(\ref{2legfibo}), one can obtain
\begin{eqnarray}
\label{golden_ratio}
\alpha^2-\alpha-1 &=&0~\implies~~
 \alpha = \frac{1+\sqrt{5}}{2}\approx 1.6180.
\end{eqnarray}
The other value of $\alpha$ obtained from the equation is negative, which is not physical.

These recursion relations have been shown to be extremely important in calculations pertaining to several important physical quantities~\cite{delgadoRecur1,fanRecur,delgadoRecur2}, including genuine multipartite entanglement \cite{hsdRecur1,hsdRecur2,sroyRecur1,sroyRecur2}.

\section{Multi-legged spin ladders}
\label{multi-legged scaling}
In this section, we begin with an analysis of the dimer covered states of the three- and four-legged RVB ladders. The main aim is to obtain a sequence for the number of coverings in the valence bond basis, similar to the Fibonacci sequence. A motivation for the study is to understand the growth of the dimer coverings when the number of legs is more than two, which is important for the asymptotic behavior of correlation and entanglement in these quantum spin ladders.\\

\emph{Three-legged ladder}--
We begin with the case for the three-legged ladder. For a $3\times N$ RVB ladder it is not possible to have odd number of rungs ($N$), as this shall result in odd number of spin sites which cannot accommodate a complete dimer covering.
The relevant Hamiltonian for the three-legged, even $N$-rung quantum spin-1/2 ladder can then be written as

\begin{eqnarray}
\mathcal{H}_\textrm{3-leg} &=&  J\sum_{m=1}^{3} \sum_{i=1}^{N-1} \vec{S}^m_i\cdot\vec{S}^m_{i+1}+ {J'}\sum_{m=1}^{2} \sum_{i=1}^{N} \vec{S}^m_i\cdot\vec{S}^{m+1}_{i} ,\nonumber\\
\label{H3}
\end{eqnarray}  
where the objects are the same as considered in Eq.~(\ref{H2}). 
We note that the Hamiltonian can not have a low-energy state with dimers along the rungs, which is radically different from the case for even-legged ladders. This is
due to the fact that there are odd number of sites along each rung (odd legs), and no complete dimer covering can form exclusively along the rungs, as in the case of the state $|1\rangle^{\otimes N}$ for the two-legged ladder. 
Consequently, the form of the recursion and the corresponding generalized sequence for the three-legged ladder are significantly different from those of the even-legged ladders.


\begin{figure}[t]
\begin{center}
\includegraphics[width=3.45in,angle=00]{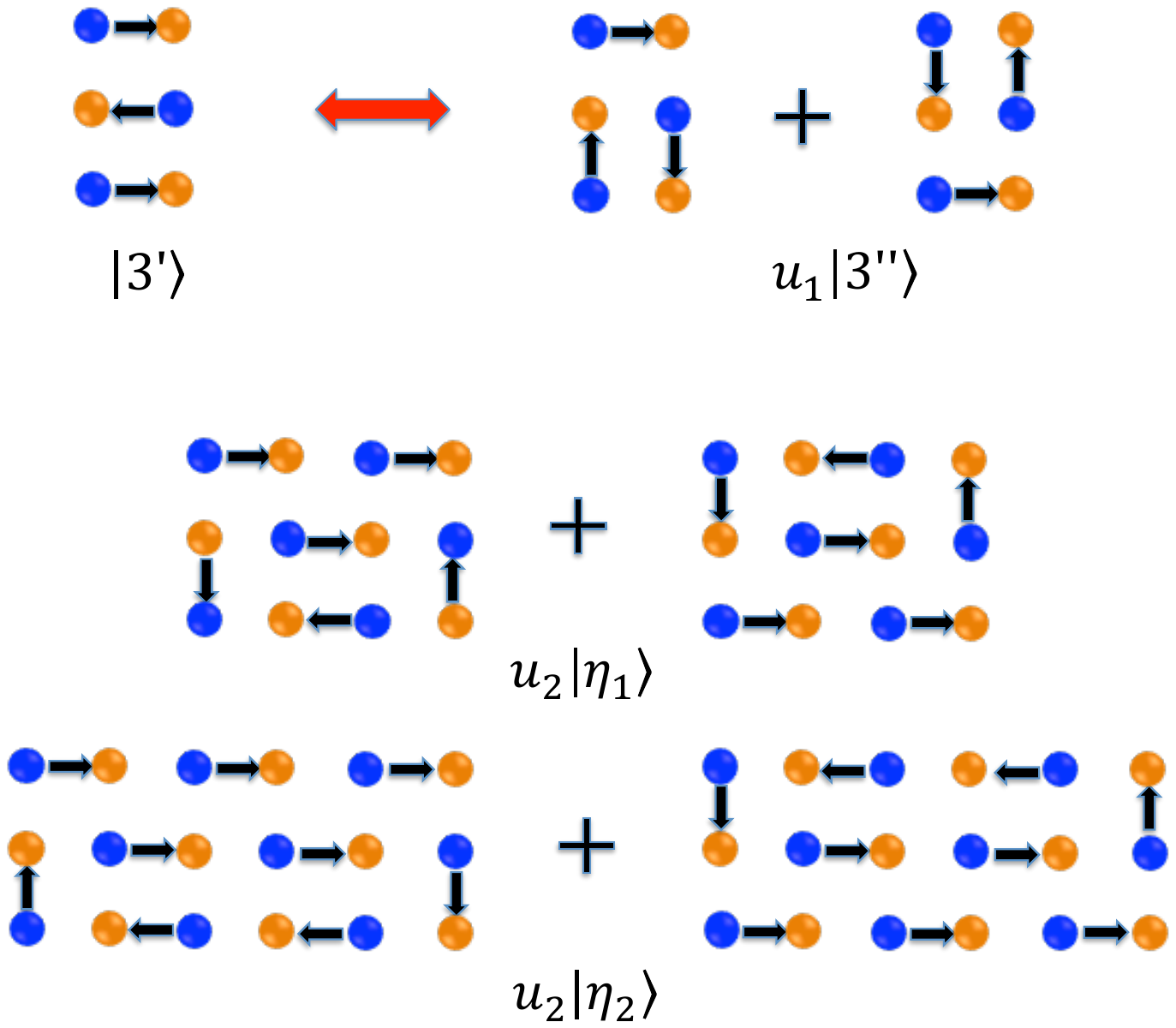}
\caption{(Color online.) The resonant dimer coverings for the three-legged quantum spin ladders. The state $|3^{''}\rangle$, is generated due to the resonance of dimers  (indicated  by two-way arrow) in $|3^{'}\rangle$. The states, $|\eta_i\rangle$ for $i > 2$, can be generated upon increasing the rungs by inserting floating horizontal singlets at the non-edge sites such that no non-singlet sharing blocks are formed. This can be observed in the above figures by looking at how $|\eta_2\rangle$ is generated from $|\eta_1\rangle$.}
\label{rvb3_terms}
\end{center}
\end{figure}
Using the RVB ansatz, for $J'$ = 0, the GS is just singlets between NN sites along the legs, with no interaction along the rungs, as given by the state $|3^{'}\rangle$ (see Fig.~\ref{rvb3_terms}). As $J'$ is turned on, the state can form new resonant states, such as $|3^{''}\rangle$, with weight $u_1 \approx J'/J$. This is similar to the resonance between the singlet pairs in the two-legged RVB ladders.  However, in the three-legged ladders, additional resonant states, $|\eta_i\rangle$, can form with weights $u_2 \approx (J'/J)^{2}$. Note here that the states $|\eta_i\rangle$'s  are configurations with $2(i+1)$ rungs, which cannot be decomposed into products of smaller blocks of states. In other words,  one can argue that these states can be considered as terms arising from a similar resonance process. Due to this fact, we assign the same coefficient to all the $|\eta_i\rangle$ states.
%
For an even $N$-rung three-legged ladder, the RVB state $|N\rangle$ can be  recursively generated (not to be confused with the $|N\rangle$ in Eq.~(\ref{2_legged_recursion}) for two-legged ladders), as given by the relation
\begin{eqnarray}
|N\rangle&=& |N-2\rangle |3^{'}\rangle+u_1 |N-2\rangle |3^{''}\rangle \nonumber\\
&+& u_2\sum_{i=1}^{N/2-1}|N-2i-2\rangle |\eta_{i}\rangle.
\label{3leg}
\end{eqnarray}
The number of dimer coverings, $Z_N$, follow the relation
\begin{eqnarray}
Z_{N}&=&Z_{\bar{3}}\times Z_{N-2}+\sum_{i=1}^{N/2-1} Z_{N-2i-2}\times Z_{\eta_{i}},\nonumber\\
&=& 3~ Z_{N-2} + 2~\sum_{i=1}^{N/2-1} Z_{N-2i-2},
\label{3_legged}
\end{eqnarray}
where, from Fig.~\ref{rvb3_terms}, we observe that $Z_{\bar{3}}$ = 3, where $|\bar{3}\rangle = |3^{'}\rangle + |3^{''}\rangle$, and $Z_{\eta_i} = 2$, $\forall~i$. 
By rearranging the terms, the above equation can further be simplified to
$
Z_{N} = 4~Z_{N-2} - Z_{N-4}.
$
We note that the same notation, $Z_{N}$, has been used for all multi-legged ladders, where $N$ is the number of rungs.

\begin{figure}[t]
\begin{center}
\includegraphics[width=3.0in,angle=00]{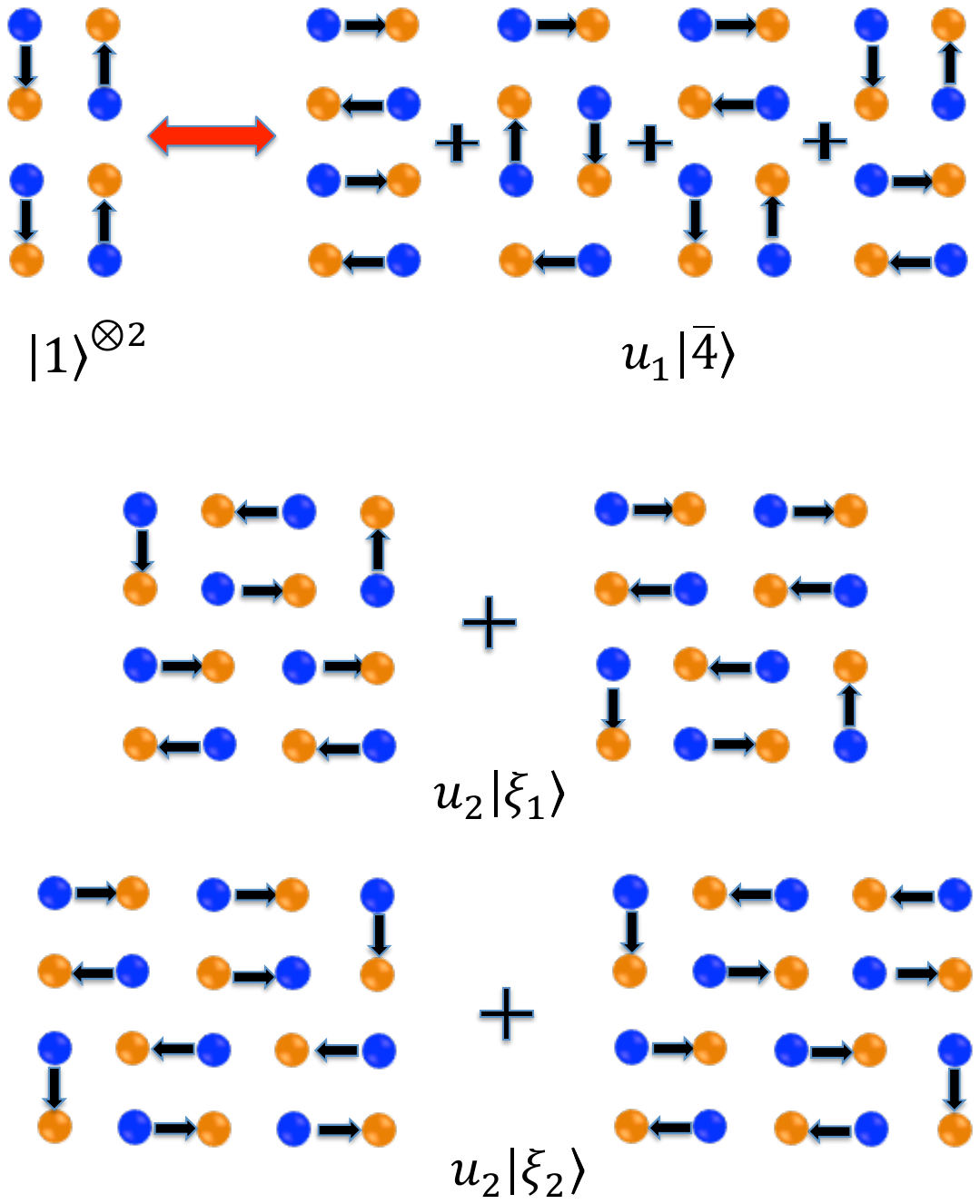}
\caption{(Color online.) The resonant dimer coverings for the four-legged quantum spin ladders. The state $|\bar{4}\rangle$, is generated due to the resonance of dimers  (indicated  by two-way arrow) in $|1\rangle^{\otimes2}$.The states, $|\xi_i\rangle$ for $i > 2$, can be generated upon increasing the rungs by inserting floating horizontal singlets at the non-edge sites such that no non-singlet sharing blocks are formed.}
\label{rvb4_terms}
\end{center}
\end{figure}

Hence, we observe that though the recursion relation for the three-legged RVB state does not yield the Fibonacci sequence, a generalized sequence for $\{Z_N\}$ can be obtained. 
The rate of divergence, at large $N$, is given by ${\alpha^{\prime}}^2=Z_{N}/Z_{N-2}$, which gives us
\begin{eqnarray}
&Z_{N} = {\alpha^{\prime}}^2 Z_{N-2}={\alpha^{\prime}}^4 Z_{N-4}&\nonumber\\
&{\alpha^{\prime}}^4-4{\alpha^{\prime}}^2+1 = 0 \implies
{\alpha^{\prime}}^2 = 2+\sqrt{3} \approx 3.7320,~~~&
\label{3legratio}
\end{eqnarray}
where the higher value of ${\alpha^\prime}^2$ (instead of $2-\sqrt{3}$) is numerically observed. The quantity ${\alpha^\prime}^2$ can be compared with the rate of divergence in the Fibonacci sequence, when one gallops over every next term in the sequence, which is equal to $\alpha^2$ = $\Phi^2$ and we find that ${\alpha^\prime}^2 > \alpha^2$. Hence, the three-legged ladder offers a higher rate of divergence when the number of rungs are increased, as compared to the two-legged ladder. If we wish to compare the number of coverings in the two- and three-legged ladders, when the number of spins are increased (and not the rungs), we must compare ${\alpha^\prime}^2$ with the number ${\alpha^3}$. Due to the even-odd disparity, the two-legged ladder seems to be more complex, in the sense of having a larger number of coverings, than the three-legged ladder, when compared with respect to the increasing number of spins. \\

\emph{Four-legged RVB ladder}-- 
The Hamiltonian for the four-legged quantum spin-1/2 ladder can be written as  \cite{delgadoRecur3}
\begin{equation}
\mathcal{H}_\textrm{4-leg} = J \sum_{m=1}^{4} \sum_{i=1}^{N-1} \vec{S}^m_i \cdot \vec{S}^m_{i+1} + J' \sum_{m=1}^{3} \sum_{i=1}^{N} \vec{S}^m_i \cdot \vec{S}^{m+1}_{i}.
\label{H4}
\end{equation}
We note that $\mathcal{H}_\textrm{4-leg}$ is similar to the case for the two-legged ladder but different from the odd-legged ladders. For $J$ = 0, the GS in the RVB ansatz is given by singlets along the rungs, $|1\rangle^{\otimes 2}$ (see Fig.~\ref{rvb4_terms}).
Subsequently, when \(J \ne 0\), but while \(J\) is still small, the resonances created are presumably those where a single or two pairs of horizontal rungs are formed, creating the state $|\bar{4}\rangle$, with relative weight  $u_1 \approx J/J^{'}$.  Additional resonant states are of the form given by $|\xi_i\rangle$.
Similarly, one can proceed to obtain the series, $\{Z_N\}$, for the four-legged RVB ladder. The $N$-rung state can again be recursively be generated as
\begin{eqnarray}
|N\rangle &=&  |N-1\rangle |1\rangle + u_1|N-2\rangle|\bar{4}\rangle + u_2\sum_{i=1}^{N-2} 
|N-i-2\rangle |\xi_i\rangle.\nonumber\\
\label{fibonacci_four}
\end{eqnarray}
Hence, the number of dimer coverings follows the recursive relation
\begin{eqnarray}
Z_{N}&=&Z_{N-1}Z_{1}+Z_{N-2}Z_{\bar{4}}
+\sum_{i=1}^{N-2}Z_{N-i-2}Z_{\xi_i},\nonumber\\
&=&Z_{N-1}+4Z_{N-2}+2\sum_{i=1}^{N-2}Z_{N-i-2},
\label{4_legged}
\end{eqnarray}
where we have inserted $Z_{1}=1$, $Z_{\bar{4}}=4$ and $Z_{\xi_i}=2, \forall~i$, and upon rearranging the terms we get
$Z_{N}=2~Z_{N-1}+3~Z_{N-2}-2~Z_{N-3}$.

As with two- and three-legged ladders, the rate of increase in the number of dimer coverings in four-legged RVB ladders, 
$\alpha^{\prime\prime}={Z_{N}}/{Z_{N-1}}$, for large $N$ is given by the relation
\begin{eqnarray} 
&Z_{N}={\alpha^{\prime\prime}}Z_{N-1}={\alpha^{\prime\prime}}^2 Z_{N-2}={\alpha^{\prime\prime}}^3
Z_{N-3},&\nonumber\\
&{\alpha^{\prime\prime}}^3-2{\alpha^{\prime\prime}}^2-3{\alpha^{\prime\prime}}+2=0~\implies~
\alpha^{\prime\prime} \approx 2.8136, &
\end{eqnarray}
where the solution of $\alpha^{\prime\prime}$, which is positive and greater than unity, has been considered. 
With respect to the number of rungs, the comparable rates of divergence for the two-, three-, and four-legged ladders are respectively $\alpha^2$, ${\alpha^\prime}^2$, and ${\alpha^{\prime\prime}}^2$. We find that the four-legged ladder has a much higher rate of divergence than the two- and three-legged ladders.
To account for the increase in coverings due to the number of legs, one can define a quantity, $\beta_l$, such that $Z_{N}^l= \beta_l^l Z_{N-2}^l$, where $l$ is the number of legs. This gives us $\beta_2$ = 2.6180, $\beta_3$ = 2.4060, and $\beta_4$ = 2.8136.
 With respect to the number of spins, for the same set of ladders,
the comparable rates are  $\alpha^6$, ${\alpha^\prime}^4$, and ${\alpha^{\prime\prime}}^3$, respectively, such that the number of spins to compare at each step are equal, i.e., $2\times 6$ =  $3\times 4$ = $4\times 3$. 
Again, the four-legged ladder has a much higher rate of divergence than the two-legged ladder, with the three-legged ladder having the lowest rate.

The recursion relations in Eqs.~(\ref{2_legged_recursion}), (\ref{3leg}), and (\ref{fibonacci_four}), gives us the closest RVB states on the two-legged, three-legged, and four-legged spin ladders, respectively, which then allows us to the calculate the number of dimer coverings in the lattice. We note that dimer arrangements on a rectangular or quadratic lattice, as is the case for the undoped spin ladders, can also be enumerated using a closed analytical relation, as given in Ref.~\cite{Kasteleyn1961}. The results obtained from our recursions agree with those obtained from the analytical relation in the asymptotic limit, with an exact match obtained for the rate of divergence in two- and three-legged ladders. However, for the doped spin lattice, no such expression to compute the coverings exist, as the arrangement of monomer-dimers on a lattice is an NP-complete problem \cite{Jerrum1987}.  In the next section, we extend our recursion relation to compute the number of dimer-hole coverings for the RVB states in doped quantum spin ladders.

\section{Doped Spin ladders}
\label{Doped RVB detail}
The doped quantum spin ladder is a  strongly-correlated electron system with an extremely rich phase structure. Many physical properties of spin lattices undergo significant changes when some doping or impurity is added to the system. In recent times, attempts have been made to establish  fundamental  connections  between  high-$T_c$ superconductivity  and  quantum  spin  fluctuations in underdoped cuprates~\cite{doped_sc},
leading to extensive study of the properties of doped RVB states~\cite{RVBdopedSup}. 

A suitable model to study a doped quantum spin ladder is the $t$-$J$ Hamiltonian  \cite{delgadoRecur2}, given by
\begin{eqnarray}
\mathcal{H}_\textrm{doped} &=& J \sum_{m=1}^{M} \sum_{i=1}^{N-1} \vec{S}^m_i \cdot \vec{S}^m_{i+1} + J' \sum_{m=1}^{M-1} \sum_{i=1}^{N} \vec{S}^m_i \cdot \vec{S}^{m+1}_{i}\nonumber\\
&-& \sum_{\langle i,j \rangle, s} t_{ij}~ \mathcal{P}_\mathcal{G} (c_{i,s}^\dag c_{j,s} + c_{j,s}^\dag c_{i,s}) \mathcal{P}_\mathcal{G}, 
\label{H_dope}
\end{eqnarray}
where $t_{ij}$ = $t,t'$, depending on whether the NN terms ($\langle i,j \rangle$) are along the legs or rungs, following the notation for $J$ and $J'$. $s$ is for the spins $\{\uparrow,\downarrow\}$ and $\mathcal{P}_\mathcal{G}$ is the Gutzwiller projection to avoid double occupancy.

\begin{figure}[h]
\begin{center}
\includegraphics[width=3.5in,angle=00]{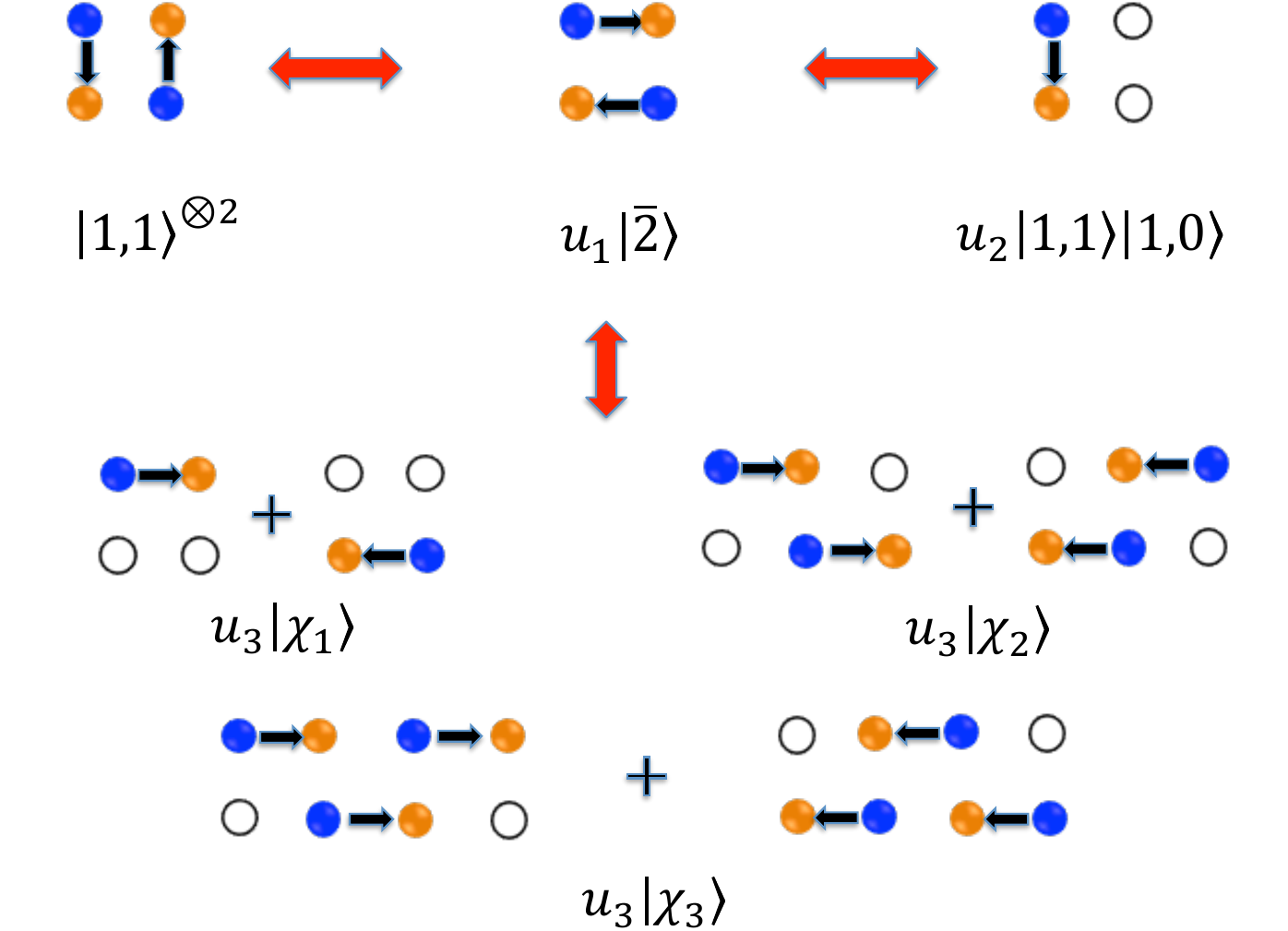}
\caption{(Color online.) The resonant hole-dimer arrangement for the doped two-legged quantum spin ladders. In the above resonant states there are two holes in the lattice. We note that $|\bar{2}\rangle$ is a possible resonant state in the absence of holes obtained from the state $|1,1\rangle^{\otimes2}$ (indicated by two-way arrow). Similarly, the states $|1,1\rangle |1,0\rangle$ and $|\chi_1\rangle$ can be obtained as a result of dimer-monomer resonance. On the other hand, the states $|\chi_i\rangle$ (for $i=2,3$) can be generated from the configuration   $|\chi_1\rangle$ by inclusion of successive horizontal dimers.}
\label{rvb2_terms_doped}
\end{center}
\end{figure}

To provide a picture for GS of the above $t$-$J$ Hamiltonian, let us consider the case for the doped two-legged, $N$-rung ladder. In the absence of doping, and in the limit $J'\gg J,t,t'$, the model reduces to the case $J=0$ in Eq.~(\ref{H2}), and the GS is given by a product of singlets along the rung, i.e., $|1\rangle^{\otimes N}$. When the system is doped with a hole, a singlet pair is broken, and the hole forms a quasi-particle associated with the unpaired spin. Introduction of a second hole replaces the unpaired electron, which is energetically favorable as compared to breaking a second singlet pair, thus forming a bound pair of holes. The resonant states that emerge upon increasing the other parameters is a superposition of hole pairs and dimers, in an effective dimer hard-core boson model \cite{delgadoRecur2}. Beyond this, additional distant hole pairs that keep the dimer arrangement complete can also be formed, as shown in Fig.~\ref{rvb2_terms_doped}.
In principle, the holes ``resonate" along with the dimers in a model containing all possible monomer-dimer arrangements. 
The doped RVB states are considered to be possible ground states of certain 1D~\cite{tj} and 2D~\cite{doped_sc2} physical models. This increases the importance of studying the physical properties related to the doped RVB states.
However, presence of dopants in the system, increases the computational complexity to levels far higher than the undoped cases, leading to fast growth of the number of coverings. Computation of quantities such as entanglement entropy becomes extremely hard even in comparatively smaller systems as compared to the undoped dimer covered states. Therefore,  estimating a generalized sequence of $\{Z_{N}\}$, as done for undoped ladders, is important and will provide valuable inputs to understand the physical properties of doped RVB ladders.

To maintain consistency with the undoped case, we investigate doped RVB ladders with two, three and four legs, for different doping percentage or concentrations. To accommodate  the doping in the system, we adopt a slightly different notation to represent the ladder states. We denote the doped RVB ladder with NN dimers as 
$|N,k\rangle$, where $N$ is the number of rungs as before, and $k$ is the number of dimers or singlets present in the system that form a complete covering. The remaining sites are vacant or are holes in the spin lattice. The doping concentration is then given by $n_{h}=1-n_e=1-\frac{2k}{l\times N}$, where $n_e$ is the singlet density, and $l$ is the number of legs. 
The valence bond basis now consists of complete dimer coverings and holes in all possible dimer-monomer arrangements in the ladder lattice, such that 
$|\phi_i^{N,k}\rangle$ = $([e_1,e_2]\otimes [e_3,e_4]\otimes\dots [e_{2k-1}e_{2k}] \otimes |0\rangle^{\otimes{h}})_i $, where $|0\rangle$ is the hole or dopant  present in $h$ vacant sites. The arrangement of $k$ singlets and $h$ holes leads to the basis states, $\{|\phi_i^{N,k}\rangle\}$, and the RVB ladder is then given  by 
$|N,k\rangle=\sum_{i=1}^{Z_{N,k}} W^\prime_i~ |\phi_i^{N,k}\rangle$.\\\\


\emph{Two-legged doped ladder}--We start from the $t$-$J$ Hamiltonian in Eq.~(\ref{doped_2leg}), and build on the hole-dimer RVB state for a two-legged quantum spin ladder.
To build on the recursion of the doped RVB state, we may consider the terms in the Eq.~(\ref{2_legged_recursion}) for the undoped two-legged RVB ladder. 
Due to presence of holes at some of the lattice sites additional resonant terms emerge in the recursion, as shown in 
Fig.~\ref{rvb2_terms_doped}. For a two-leg, $N$-rung doped RVB ladder with $k$ dimers, the recursion relation reads as
\begin{widetext}
\begin{eqnarray}
\label{recur_doped_2leg}
|N,k \rangle &=&|N-1,k-1\rangle|1,1\rangle + u_1 |N-2,k-2\rangle |\bar{2}\rangle+  u_2 |N-1,k\rangle |1,0\rangle+u_3\sum_{i=1}^k |N-i-1,k-i\rangle |\chi_i\rangle.\nonumber\\
\end{eqnarray}

%
%
Using this, we find that the number of dimer coverings in $|N,k\rangle$ is given by
\begin{eqnarray}
Z_{N,k}&=&Z_{N-1,k-1}Z_{1,1}+Z_{N-2,k-2}Z_{\bar{2}}+Z_{N-1,k}Z_{1,0}+\sum_i Z_{N-i-1,k-i}Z_{\chi_i}.\nonumber\\
&=&Z_{N-1,k-1}+Z_{N-2,k-2}+Z_{N-1,k}+2\sum_i Z_{N-i-1,k-i},
\label{doped_2leg}
\end{eqnarray}
\end{widetext}
where we have plugged the terms $Z_{1,1}=1$, $Z_{\bar{2}}=1$ and $Z_{\chi_i}=2$, and  rearranged the terms to obtain
$Z_{N,k} = 2Z_{N-1,k-1}+Z_{N-1,k}+Z_{N-2,k-1}-Z_{N-3,k-3}$.\\

\begin{figure}[h]
\includegraphics[width=8cm]{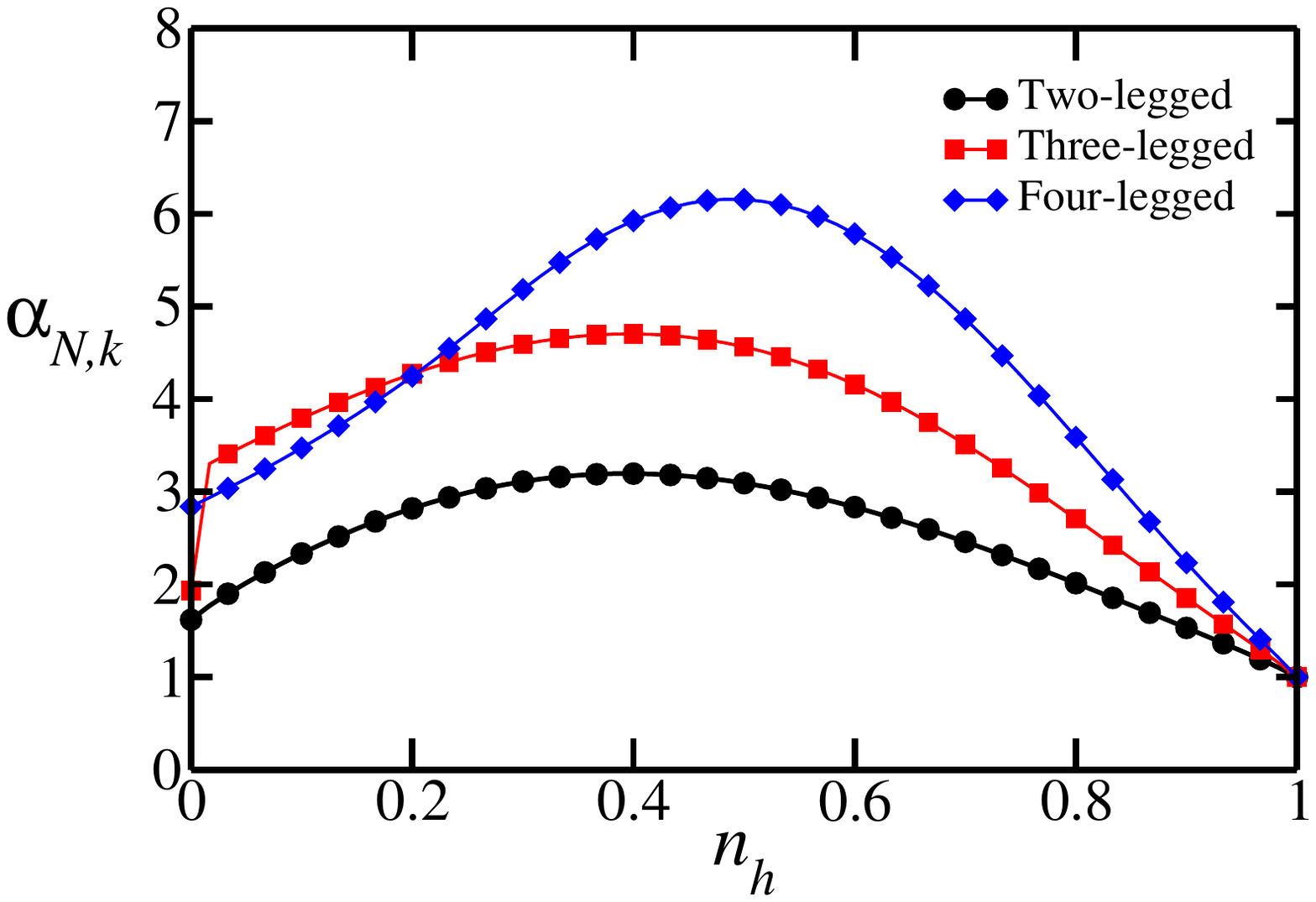}
\caption{(Color online.) The changing face of the generalized sequences with increase in dopants. We exhibit here the
rate of increase of the number of coverings $Z_{N,k}$, given by $\alpha_{N,k}$, for different hole concentration $n_{h}$.
For each doped RVB ladder, with a different number of legs, there is a particular value of hole concentration, $n_{h_c}$, where $\alpha_{N,k}$ is maximum. 
All quantities are dimensionless. We note that there is a sharp increase in  $\alpha_{N,k}$ for finite doping in the three-legged ladder. This is due to the fact that doping allows formation of extra coverings in the recursion relation in Eq.~(\ref{recur_doped_3leg}) as compared to the undoped case in Eq.~(\ref{3leg}).}
\label{doped_ladder_ratio}
\end{figure}

Thus we obtain a recursion relation to generate the series $\{Z_{N,k}\}$. From Eq.~(\ref{doped_2leg}), it is clear that the number of dimer coverings has a dependence on the hole concentration ($n_h$). 
Similarly, the rate of divergence ($\alpha_{N,k}$) is a function of $n_h$, and though a closed analytical form is cumbersome to present, the function can be numerically estimated using the recursion with relative ease.
In Fig.~\ref{doped_ladder_ratio}, we plot  $\alpha_{N,k}={Z_{N,k}}/{Z_{N-1,k}}$, for varying hole concentration ($n_h$)  (solid black circles), where $k'=k-Nh$. We observe that the  effect of doping  in the two-legged RVB ladder causes $\alpha_{N,k}$ to deviate from golden ratio. Starting off from the golden ratio at $n_h$ = 0, for low  hole concentration, the quantity   $\alpha_{N,k}$ increases with  $n_h$  and reaches its maximum value  at $n_{h_c}\approx 0.44$.  Further increase of hole concentration reduces the rate of divergence. \\


\emph{Three-legged doped ladder}--
For the three-legged ladder, the hole-dimer RVB state emerging from a possible $t$-$J$ model, with a large number of spins, can be constructed from smaller system sizes, following the recursion
\begin{widetext}
\begin{eqnarray}
|N,k\rangle&=&|N-1,k\rangle |1,1\rangle+ u_1 |N-1,k\rangle|1,0\rangle+ u_2\sum_{i=1}^{{N}/{2}}\sum_{j=0}^{3i-1} |N-2i,k-3i+j\rangle |\eta_i^j\rangle,
\label{recur_doped_3leg}
\end{eqnarray}
where  $|\eta_{i}^j\rangle$'s can be derived directly from the terms $|\eta_i\rangle$ in Eq.~(\ref{3leg}) (shown in Fig.~\ref{rvb3_terms}) by introducing a pair of holes. 
Hence, the number of hole-dimer coverings follows the relation
\begin{eqnarray}
Z_{N,k}&=&Z_{N-1,k-1}Z_{1,1}+Z_{N-1,k}Z_{1,0}+\sum_{i=1}^{{N}/{2}}\sum_{j=0}^{3i-1}Z_{N-2i,k-3i+j} Z_{\eta_i}^j,
=2 Z_{N-1,k-1}+Z_{N-1,k}
+3\sum_{i=1}^{{N}/{2}}\sum_{j=0}^{3i-1}Z_{N-2i,k-3i+j},\nonumber\\
\label{doped_3leg}
\end{eqnarray}
\end{widetext}
where we have used, $Z_{1,1} = 2$, $Z_{1,0}=1$ and $Z_{\eta_i^j}=3$. 
For the three-legged doped RVB ladder, the rate of divergence $\alpha_{N,k}$  varies with the hole concentration $(n_h )$ in a similar fashion to that observed in case  of the  doped two-legged ladder (see the red squares in Fig.~\ref{doped_ladder_ratio}). However, here  $\alpha_{N,k}$ reaches its maximum value  at a lower hole density ($n_{h_c}\approx 0.38$). In addition to this, for an arbitrary doping concentration,  the value of the function $\alpha_{N,k}$  remains higher than  that of the value obtained for the  two-legged case. \\

\emph{Four-legged doped ladder}--
We now derive the recursion relation for a doped four-legged RVB ladder for arbitrary number of rungs ($N$) and arbitrary hole concentration ($n_h$). 
The recursion relation for a four-legged $N$-rung ladder is given by
\begin{widetext}
\begin{eqnarray}
|N,k\rangle&=& |N-1,k-2\rangle |1,2\rangle+u_1|N-1,k\rangle |1,0\rangle+ u_2 |N-1,k-1\rangle|1,1\rangle\nonumber \\&+& u_3\sum_{i=1}^{{N}/{2}}\sum_{j=0}^{4i-1} |N-2i,k-4i+j\rangle |\xi^i_j\rangle,\nonumber\\
\label{recur_doped_4leg}
\end{eqnarray}
where $|\xi_i^j\rangle$ can  directly be derived from the terms $|\xi_i\rangle$ given in Eq.~(\ref{4_legged}) (shown in Fig.~\ref{rvb4_terms})   by introducing pair of holes.
Hence, the number of  dimer coverings follow the recursion
\begin{eqnarray}
Z_{N,k} &=& Z_{N-1,k-2}Z_{1,2}+Z_{N-1,k-1}Z_{1,1}+Z_{N-1,k}Z_{1,0}+\sum_{i=1}^{N}\sum_{j=0}^{4i-1}Z_{N-2i,k-4i+j}Z_{\xi_i^j},\nonumber\\
 &=&Z_{N-1,k-2}+3Z_{N-1,k-1}+Z_{N-1,k}+2\sum_{i=1}^{N}\sum_{j=0}^{4i-1}Z_{N-2i,k-4i+j}.
\label{doped_4leg}
\end{eqnarray}
\end{widetext}
with $Z_{1,2}=1$, $Z_{1,1}=3$, $Z_{1,0}=1$ and $Z_{\xi^j_i}=2$, which gives us the sequence $\{Z_{N,k}\}$.
The rate of divergence, $\alpha_{N,k}$, as a function of $n_h$, is plotted in Fig.~\ref{doped_ladder_ratio}.
In this case, for $0<n_h\lesssim 0.247$, the quantity   $\alpha_{N,k}$  has a value intermediate between the two- and three-legged doped ladders. Moreover,  $\alpha_{N,k}$ reaches its maximum value, at a hole concentration  ($n_{h_c}\approx 0.5$), which is much higher than that of the previous cases for doped ladders. Further increase of the hole concentration leads to reduction of the diverging rate.

\section{Valence Bond Entanglement Entropy in quantum spin ladders}
\label{VBE}

While several significant physical quantities have been explored for RVB ladders, including spin correlation functions \cite{delgadoRecur1,fanRecur,delgadoRecur2} and entanglement \cite{hsdRecur1,hsdRecur2,sroyRecur1,sroyRecur2}, most of these works rely on employing numerical approaches based on tensor-networks \cite{expo1,scala}, quantum Monte-Carlo \cite{wiese}, mean-field \cite{gopalan}, or sophisticated recursion methods \cite{hsdRecur1,hsdRecur2,sroyRecur1,sroyRecur2}.
In this section, we show how the obtained generalized sequence of dimer coverings allows us to directly obtain some interesting physical properties of the quantum spin ladders, without resorting to any complex numerical method. 
The physical property that we investigate is the valence bond entanglement entropy \cite{aletVBE,wojcikVBE}, which provides an alternative approach to estimate the scaling behavior of entanglement entropy in quantum spin systems with SU(2) symmetry. The scaling behavior of entanglement entropy in the  ground state of strongly correlated systems has been widely researched in recent years to study area law \cite{sredAL}, critical phenomena \cite{vidalEE,latorreEE} and ground state topology \cite{wenTopo,kitaevTopo} (also see, \cite{fazioMB,eisertAL,USAdv}). Although intrinsic to the characteristics of a physical system, entanglement entropy is not easily computable in systems.
To obtain the scaling behavior, it is imperative to consider larger systems beyond the realm of exact diagonalization. Moreover, tensor network approaches such as density matrix renormalization group~\cite{whiteDMRG} and matrix product states~\cite{MPS} seem unsuitable in dimensions higher than one.

%

\begin{figure}[h]
\begin{center}
\includegraphics[width=8cm]{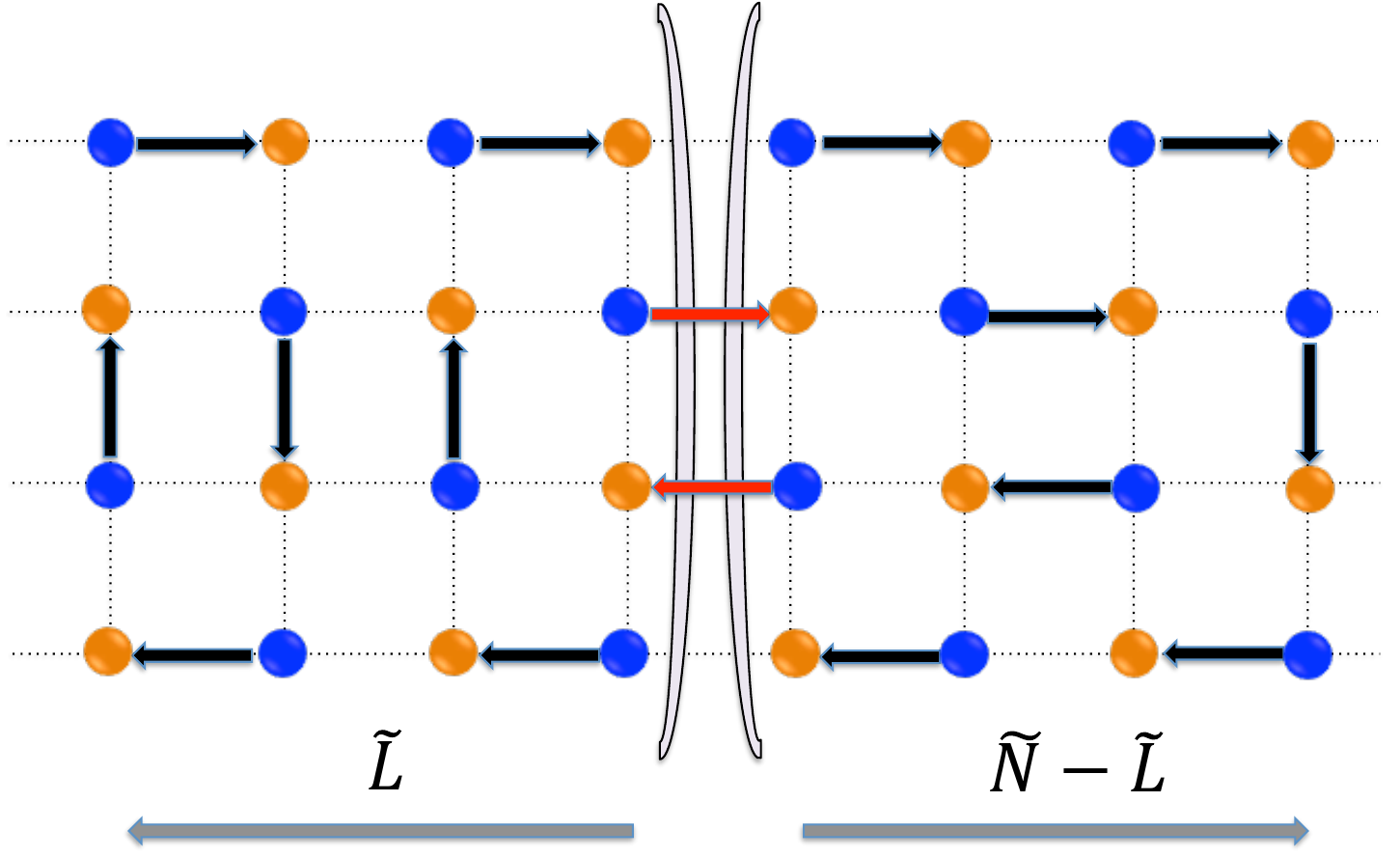}
\end{center}
\caption{(Color online.) A dimer covered quantum spin ladder with a bipartition across $\tilde{L}:\tilde{N}-\tilde{L}$ spins.  The covering contains $a_i$ = 2 dimers across the boundary, shared between the blocks of  $\tilde{L}$ and $\tilde{N}-\tilde{L}$ spins, which contributes to the valence bond entanglement entropy in Eq.~(\ref{eqVBE}). The above dimer arrangement constitutes a single covering in the valence bond basis 
$\{|\phi_i\rangle\}$. For any given ladder, $Z_N$ such coverings are possible, each with $a_i$ singlets present across the boundary. }
\label{vbe}
\end{figure}

The valence bond entanglement entropy in quantum spin systems is defined as the average number of dimers in the valence bond basis that cross the boundary between two bipartitions. Though not directly comparable, the  valence bond entanglement entropy has been used to effectively predict the scaling properties of entanglement entropy in the different phases of the nonfrustrated 2D spin-1/2 Heisenberg model \cite{aletVBE,wojcikVBE,melkoVBE,VBE}, using Quantum Monte Carlo simulations in the valence bond basis \cite{sandvikVB}. However, the predicted corrections to the area law at criticality does not always seem to be
comparable \cite{melkoVBE}. To define the valence bond entanglement entropy, one looks for possible representations of a quantum spin state with SU(2) symmetry in the overcomplete dimer-covered valence bond basis $\{|\phi_i\rangle\}$ \cite{aletVBE,wojcikVBE}. For instance, 
the total spin singlet ground state, of a quantum spin-1/2 Heisenberg system, with even number of spins ($\tilde{N}$) can be represented as
$
|\Psi\rangle_{\tilde{N}} = \sum W_i |\phi_i\rangle$,
where $W_i$ are the 
coefficients~\cite{VBE1,sandvikVB,marshall} for each valence bond covering. Let us consider the state, 
$|\Psi\rangle_{\tilde{N}}$, in a bipartition given by $\tilde{L}:\tilde{N}$-$\tilde{L}$, such that for the basis state $|\phi_i\rangle$, $a_i$ singlets are shared between $\tilde{L}$ and $\tilde{N}$-$\tilde{L}$, i.e., $a_i$ singlets cross the boundary  (see Fig.~\ref{vbe}). The valence bond entropy is then defined as \cite{aletVBE}
%
%

\begin{eqnarray}
S_{L}= \frac{\ln 2}{\sum |W_i|}\sum_i a_i |W_i|,
\label{eqVBE}
\end{eqnarray}
where, for consistency with later sections, $S_L$ is defined for the number of rungs $L$ in the ladder, such that $\tilde{L}$ = $l \times L$ ($l$ is the number of legs in the multi-legged ladder). For one dimensional system, $L$ = $\tilde{L}$. 
For doped spin lattices, each  basis state in the overcomplete $\{|\phi_i\rangle\}$ will now represent a complete hole-dimer covering, and  the above definition for valence bond entropy is then used in a similar fashion.

\begin{figure}[h]
\includegraphics[width=8cm]{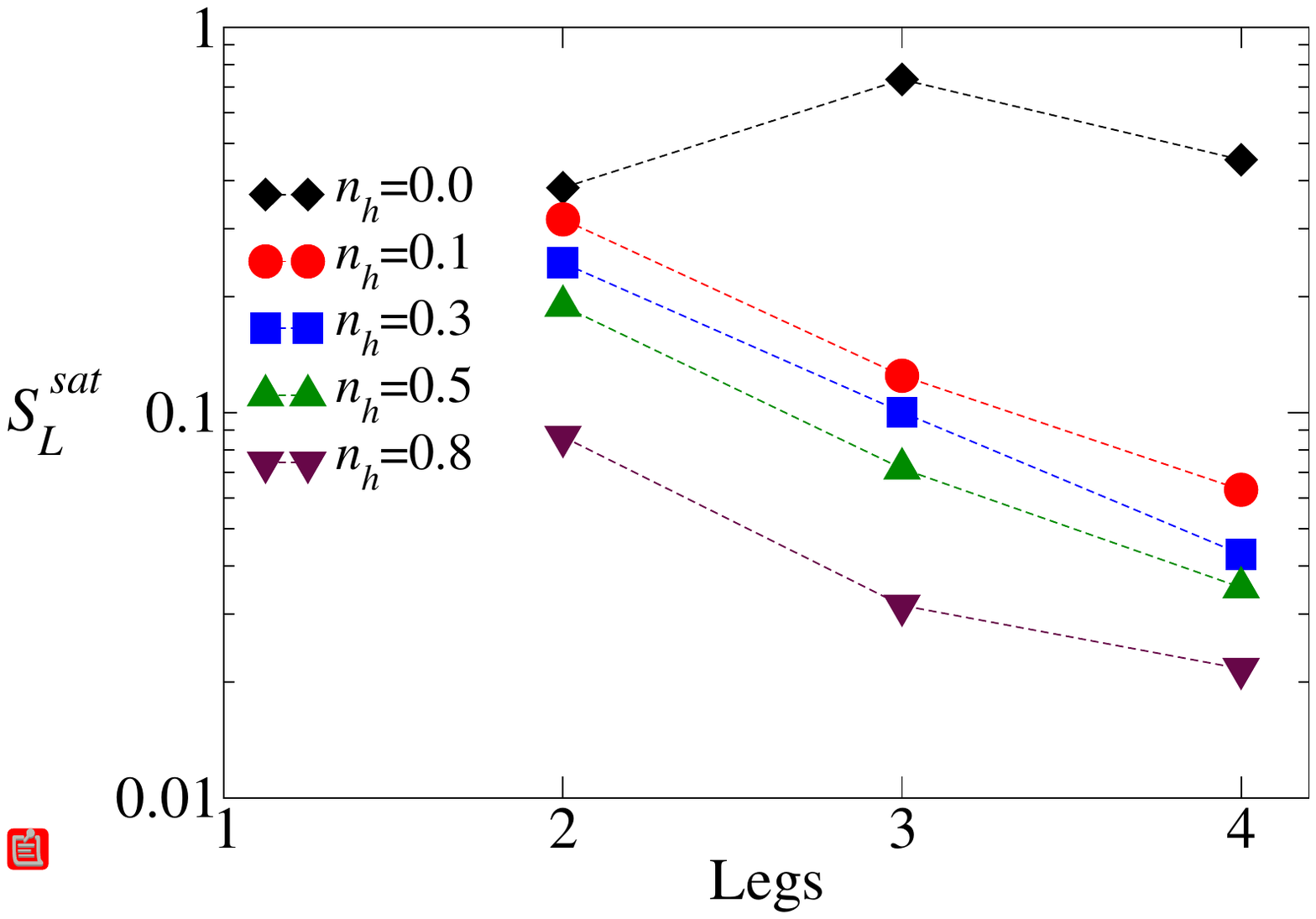}
\caption{(Color online.) Variation of saturated valence bond entanglement entropy with the number of legs. We consider quantum spin ladders, with $N$ = 100 rungs ($200-400$ spins), for different doping concentrations, $n_h$, and investigate the behavior of $S^{sat}_L$ with increasing number of legs. The undoped RVB ladder corresponds to the case where $n_h$ = 0. The vertical axis is in ebits and is plotted on a logarithmic-scale.
}
\label{figure1}
\end{figure}

\begin{figure}[h]
\includegraphics[width=8cm]{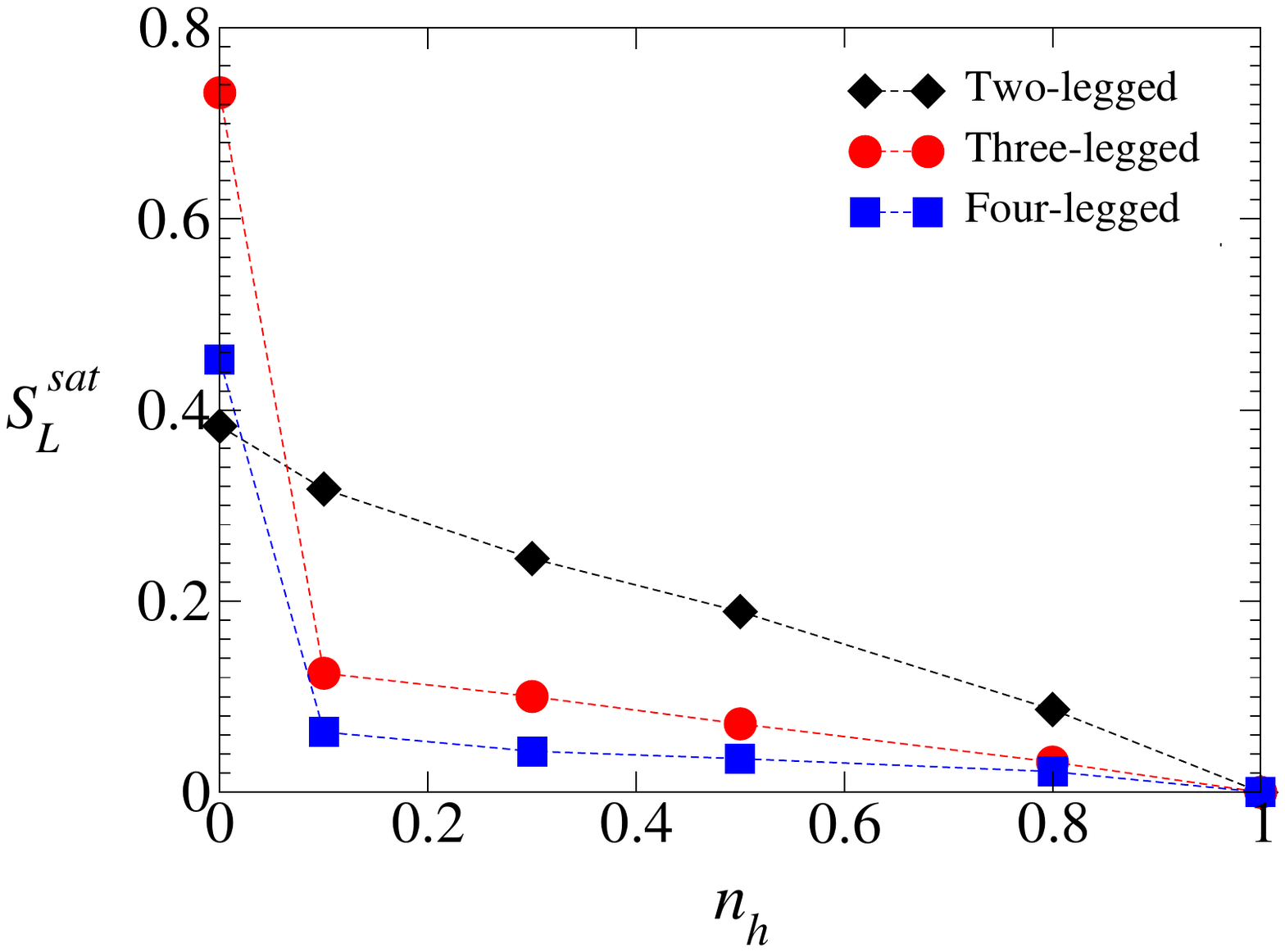}
\caption{(Color online.) Variation of saturated valence bond entanglement entropy with the doping concentration. We consider quantum spin ladders, with $N$ = 100 rungs, for different numbers of legs, and investigate the behavior of $S^{sat}_L$ with increasing doping concentration. The undoped case corresponds to the case where $n_h$ = 0, with the completely vacant lattice given by  $n_h$ = 1. The vertical axis gives us the variation of $S_L^{sat}$ with the numbers of legs, for fixed $n_h$. The information presented in this figure therefore complements those in Fig. 8. The vertical axis is ebits, while the horizontal one is dimensionless.}
\label{figure2}
\end{figure}

We now look at how the sequence $\{Z_N\}$ can be used to estimate the valence bond entropy across an arbitrary boundary, for a two-legged RVB ladder with $N$ rungs. For simplicity, let us consider the case where $J = J'$ in Eq.~(\ref{H2}) for the two-legged ladder, which leads to uniform resonance, $u_1 = 1$, such that all resonant states are equally weighted. This corresponds to $W_i = 1, \forall~i$, in Eq.~(\ref{eqVBE}). For the boundary between the ${L}$ and $L+1$ rungs, such that the ladder is divided into ${L}$ and $N$-$L$ rungs, the only coverings that contain a singlet pair across the boundary between the ${L}$ and $L+1$ rungs, are those that have the state $|\bar{2}\rangle$ at these rungs. Such coverings come from the state, $|L-1\rangle\otimes|\bar{2}\rangle_{L,L+1}\otimes|N-L-1\rangle$. Therefore the total number of singlets is $\sum_i a_i$ = $2 Z_{L-1}\times Z_{N-L-1}$, and the valence bond entanglement entropy is given by
\begin{equation}
S_L = \ln 2~ \frac{\sum a_i}{\sum W_i} =  \frac{2 Z_{L-1} Z_{N-L-1}}{Z_N}\ln2, 
\label{VBE2}
\end{equation}
as $\sum_i W_i = Z_N$, since $W_i = 1~\forall~i$. 
%
Since, we already know the sequence for $\{Z_N\}$, the above quantity can be calculated for arbitrary large number of spins in the ladder. In our study, we obtain the asymptotic or saturated value of $S_L$, by considering, but not limited to, $N$ = 100 rungs in Fig.~\ref{figure1} and Fig.~\ref{figure2}. In fact, $S_L$ saturates at a relatively low value of ${L}$ ($\sim 10$) rungs. The behavior shows that the valence bond entanglement entropy follows the area law. This is consistent with exponentially decaying correlations in two-legged RVB ladders \cite{expo,expo1}, which suggests an area law for the entanglement entropy \cite{decay_AL}. Therefore, in this case, the scaling of $S_L$ emulates the behavior of entanglement entropy.
The saturation value of $S_L$ can be expressed in terms of the rate of divergence, $\alpha$. For ${L}$ = $N/2$ and $S_L^{sat}$ = $S_{N/2}$, in Eq.~(\ref{VBE2}), we have
\begin{eqnarray}
S_L^{sat}&=&  \frac{2  Z_{N/2-1} \times Z_{N/2-1}}{Z_{N}}\ln 2
=\frac{2 Z_{N/2}}{\alpha^{N/2+2}} \ln 2
= \frac{2 p}{\alpha^2}\ln 2,\nonumber\\
\end{eqnarray}
where, from Eq.~(\ref{2-legged_fibonacci}), we have used that $Z_{x+y} = \alpha^{y} Z_{x}$ for positive integers and large $x$ and $y$. Here $p = {Z_{N/2}}/{\alpha^{N/2}}$ is a constant for large $N$. 


Along similar lines, one can obtain the valence bond entanglement entropy in both doped and undoped multi-legged RVB ladders, using just the generalized sequence $\{Z_N\}$ derived in the previous sections.
The quantity $S_L$ for the three-legged ladder, bipartitioned across $L:N$-$L$ rungs, can be obtained from the recursion in Eq.~(\ref{3_legged}), by enumerating the singlets crossing the boundary. The expression of $S_L$ is then given by
\begin{eqnarray}
S_{L} &=& \frac{\ln 2}{Z_N}~ (5 Z_{L-1} \times Z_{N-L-1}\nonumber \\
&+&2\sum_{i=1}^{N-L-1} Z_{L-2}\times Z_{N-L-1-i}).
\label{vbe_3_leg}
\end{eqnarray}
To obtain $S^{sat}_L$, we look at the leading terms in Eq.~(\ref{vbe_3_leg}) for the partition $L = N/2$. This gives us
\begin{eqnarray}
S_L^{sat} &=& \frac{5 \ln2}{Z_N}~Z_{\frac{N}{2}-1}^2 (1+\frac{2}{5\alpha^{\prime 2}} + \frac{2}{5\alpha^{\prime 3}}+\dots),\nonumber\\
&\approx& {5 \ln2}\times \frac{Z_{N/2}}{\alpha^{\prime {N}/{2}+2}} = \frac{5 \ln 2~ p^\prime}{\alpha^{\prime 2}}, 
\end{eqnarray}
where $p^\prime = {Z_{N/2}}/{\alpha^{\prime N/2}}$, at large $N$. From Fig.~\ref{figure1}, we observe that $S^{sat}_L$ here is higher than that for the two-legged RVB ladder. Therefore, an increase in the number of legs increases the average number of singlets across the block boundary, when compared on the axis of the number of rungs. One can make the same comparison on the axis of the number of spins, instead of rungs, 
the saturation values would remain unaltered.

For the $N$-rung four-legged RVB ladder, the expression for the valence bond entanglement entropy is
\begin{eqnarray}
S_L &=&  \frac{\ln 2}{Z_N}~(10~ Z_{L-1}\times Z_{N-L-1}\nonumber\\
&+&2\sum_{i=1}^{N-L-3} Z_{L-1}\times Z_{N-L-3-i}),
\label{vbe_4_leg}
\end{eqnarray}
where the contributions to valence bond entanglement entropy only arise from the $|\bar{4}\rangle$ and $|\xi_i\rangle$ terms in Eq.~(\ref{3leg}). 
Similar to the case for three-legged ladders, only the first term in Eq.~(\ref{vbe_4_leg}) dominates while contributing to $S_L$, and hence $S^{sat}_L$ can be written as functions of the diverging rate as 
${10 p^{\prime\prime} }/{\alpha^{\prime\prime 2}}$, where $p^{\prime\prime} = {Z_{N/2}}/{\alpha^{\prime\prime N/2}}$, at large $N$. Interestingly, in this case (see Fig.~\ref{figure1}), the value of the saturation entropy, $S_{sat}$, exceeds the value  obtained for the two-legged case. However, it is lower than that of the value of $S_{sat}$, for three-legged  undoped ladder. This is due to the fact that unlike the three-legged case, here the number of dimer coverings those contribute to valence bond entanglement entropy are less in number due to the difference in odd-even RVB state recursion.

The valence bond entanglement entropy is also computable for the doped RVB ladders provided the generalized sequence $\{Z_N\}$ and the corresponding divergence rate are estimated, as obtained from Eqs.~(\ref{recur_doped_2leg})-(\ref{doped_4leg}) in Sec.~\ref{Doped RVB detail}.
However, there is a  considerable increase in the inherent complexity in the system, due to the more complicated dimer-monomer arrangement in the bipartite lattice. The expression for $S_L$, in terms of the generalized sequences and fixed $n_h$, for the two-, three-, and four-legged doped RVB ladders are shown below:
\begin{widetext}
\begin{eqnarray}
S_{L,n_h}^{2-\textrm{leg}} &=&  \frac{\ln 2}{Z_{N,k}}\times(2\sum_{i=0}^{k-2} Z_{L-1,k-2-i}Z_{N-L-1,i} + 2\sum_{i=1}^{N-L} \sum_{j=0}^{k-i}Z_{N-L-i,k-i-j}Z_{L-1,j}),\label{doped_1}\\
S_{L,n_h}^{3-\textrm{leg}} &=&  \frac{\ln 2}{Z_{N,k}}\times(2\sum_{i=0}^{k-2} Z_{L-1,k-2-i}Z_{N-L-1,i} + 2\sum_{i=1}^{N-L} \sum_{j=0}^{k-i}Z_{N-L-i,k-i-j}Z_{L-1,j}),\label{doped_2}\\
S_{L,n_h}^{4-\textrm{leg}} &=& \frac{\ln 2}{Z_{N,k}}\times (\sum_{j=1}^{3}j \sum_{i=0}^{k-j} Z_{L-1,k-j-i}Z_{N-1-L,i}+4\sum_{i=1}^{{N}/{4}}~\sum_{j=0}^{6i-3}Z_{N-4i-L-1,k-6i-j}Z_{L-1,j}).\label{doped_3}
\end{eqnarray}
\end{widetext}
The saturated value of the valence bond entanglement entropy, $S^{sat}_{L}$, can be estimated for different values of the hole concentration, $n_h$, using Eqs.~(\ref{doped_1})-(\ref{doped_3}). To avoid furnishing elaborate algebra, we do not present the expression for  $S^{sat}_{L}$. However the values of the same, for doped ladders with different number of legs and doping concentration, have been plotted in Figs.~\ref{figure1} and \ref{figure2}.

In Fig.~\ref{figure1}, we look at the variation of $S^{sat}_{L}$ in quantum spin ladders with increasing number of legs, for different doping concentrations $n_h$. The undoped RVB ladder, given by $n_h$ = 0, corresponds to the generalized sequences obtained in Secs.~\ref{fibonacci_summary} and \ref{multi-legged scaling}, and are not the limiting cases for the doped spin ladders. The figure shows a distinct contrast between the $S^{sat}_{L}$ in doped and undoped RVB ladders, for different number of legs. Specifically, for $n_h$ = 0, there exists a sharp difference between odd and even-legged ladders. Thus, in the absence of doping, there is an odd-even dichotomy, which arises from their respective spin Hamiltonians. For odd-legged spin ladders there is a lack of spin-singlets exclusively along the rungs in the recursion, such as the states $|1\rangle^{\otimes{N}}$ in Figs.~\ref{rvb2_terms} and \ref{rvb4_terms}, as there are odd number of sites along a rung. These states increase $Z_N$, without contributing to $S_L$ ($a_i = 0$). Hence, in the absence of these states in odd-legged ladders, the average number of singlets crossing any $L:N$-$L$ boundary increases, and the $S^{sat}_{L}$ is considerably higher. In contrast, in the $t$-$J$ Hamiltonian for doped quantum spin ladders, with a monomer-dimer model, odd- and even-legged ladders are treated on a more equal footing, resulting in a monotonic decrease of $S^{sat}_{L}$ with increasing number of legs. 
This is evident from the fact that rung-singlets in odd-legged ladders, which are missing in the undoped case, instantly become available in the presence of dopants or holes at the odd site of the rung. Hence, in terms of the valence bond entanglement entropy, the odd-even disparity between quantum spin ladders is mitigated due to doping in the system.

Subsequently, in Fig.~\ref{figure2}, we look at the variation of $S^{sat}_{L}$ with increasing doping concentration, for different multi-legged quantum spin ladders. We observe that introduction of even a small doping in higher-legged ladders (legs $> 2$), has a strong effect on $S^{sat}_{L}$. This is due to the fact that the transition from a multi-legged dimer model, arising from a Heisenberg spin Hamiltonian, to a dimer hard-core boson or a monomer-dimer model coming from a $t$-$J$ Hamiltonian, in the case of doping, drastically changes the nature of dimers formed in the system, and subsequent recursion relations, especially in higher-legged ladders. In other words, the transition from spin dimer to monomer-dimer arrangements strongly affects the generalized sequences $\{Z_N\}$, which give us $S^{sat}_{L}$. The presence of dopants in the system allow for hole-dimer coverings which are not only radically different but are altogether absent from the undoped case. Although, the saturated value changes continuously with varying doping concentration, there is a sharp transition when even a small, finite, doping is introduced in the RVB ladder.
Interestingly, if we look along the vertical axis, and compare the plots with Fig.~\ref{figure1}, we observe that $S^{sat}_{L}$ increases with the number of legs in the undoped case, in contrast to doped spin ladders. The possible explanation for the finding is that for the same doping concentration, the number of holes in the ladder is significantly larger in higher-legged ladders leading to significantly lower saturated values of entanglement entropy across any bipartition, as compared to the undoped case, where higher fraction of singlet contributions exist for a  higher number of legs. This can be explained  if we look at the recursion relations of the RVB states for doped multi-legged ladders, in Eqs.~(\ref{recur_doped_2leg}), (\ref{recur_doped_3leg}), and (\ref{recur_doped_4leg}). 
We observe that the bulk of the terms in the recursion are generated from smaller configurations. Hence, the number of terms which are product across any bipartition grow more quickly with increasing number of legs as compared to the terms that cross the bipartition and contribute to the VBE. This affects the overall fraction of terms which contributes to VBE, leading to the lowering of the quantity with an increasing number of legs. Therefore, ladders with lesser number of legs, turn out to be more robust against the effect of doping or impurity in the system. The two-legged ladder serves as an intermediary case between undoped and doped multi-legged quantum spin ladders, under the RVB ansatz.

\section{Conclusions}
\label{Discussion}
In this work, we show how a  sequence enumerating the number of valence bond coverings in both doped and undoped multi-legged quantum spin ladders can be estimated for arbitrary number of spins. Starting from well-motivated spin models, arising from the Heisenberg and $t$-$J$ Hamiltonians, we describe the low-lying resonant ground states, under the resonating valence bond ansatz, which provides us with the recursion relation for the RVB ladder state. Building on the Fibonacci sequence for two-legged RVB ladders, we analytically construct such generalized sequences for higher-legged ladders, under different doping concentrations. Importantly, the derived sequences can be used to directly calculate important physical quantities without employing any sophisticated numerical methods.

The valence bond entanglement entropy in these quantum spin systems with SU(2) symmetry can be obtained as a direct consequence of derivation of these generalized sequences. We derive the analytical expression for the valence bond entanglement entropy in terms of the generalized sequences. We find the scaling of valence bond entanglement entropy in these systems saturates to a finite value, upon variation of the block-size of the ladders, which is typical of the behavior of entanglement entropy in noncritical low-dimensional quantum spin systems where the area law holds. By observing the asymptotic or saturated value of the valence bond entanglement entropy, in both doped and undoped RVB ladders, we investigate the effects of doping on the quantum spin ladders. In particular, we observe a distinct dichotomy between undoped odd- and even-legged ladders through the fact that the three-legged ladder has a higher saturation value of valence bond entanglement entropy compared to the even-legged ladders. This is expected from the very different recursion relations, for the respective states. The absence of complete rung-singlets, with null contributions to the measure, increase the average valence bond entanglement entropy in odd-legged ladders. In contrast, upon doping the quantum spin ladder, the odd-even dichotomy is mitigated as the $t$-$J$ interactions, and the subsequent monomer-dimer arrangements in the recursion relation, do not differentiate between odd- and even-legged ladders. For instance, the absence of rung-singlets in odd-legged ladders, is reversed by presence of dopants in the spin ladder. Thus, as shown in our work, the presence of dopants in the quantum spin ladders appears to mitigate the odd-even dichotomy in the scaling of valence bond entanglement entropy.

The derivation of sequences, akin to the Fibonacci sequence, to obtain the total number of
dimer or dimer-monomer coverings in the valence bond basis for quantum spin systems
with SU(2) symmetry may prove to be useful in several other settings, especially in
complex- or higher-dimensional strongly-correlated models, where numerical approaches
based on tensor-network or quantum Monte-Carlo may not be easily accessible. Apart from
investigating cooperative phenomena in such systems, such investigations could also be
applied to estimate important physical quantities in these systems, including entanglement,
correlation functions, and other information-theoretic system parameters and characteristics that are experimentally and computationally useful.

\section{Acknowledgments}
The research of SSR was supported in part by the INFOSYS scholarship for senior students.
HSD acknowledges funding by the Austrian Science Fund (FWF), project no.~M 2022-N27, under the Lise Meitner programme of the FWF.

\end{document}